# New Jarlskog determinant from Physics above the GUT Scale

Bipin Singh Koranga and S. Uma Sankar

*Department of Physics, Indian Institute of Technology Bombay, Mumbai 400076, India.*

## Abstract

We study the Planck scale effects on Jarlskog determinant. Quantum gravitational (Planck scale) effects lead to an effective $SU(2)_L \times U(1)$ invariant dimension-5 Lagrangian involving neutrino and Higgs fields, which gives rise to additional terms in neutrino mass matrix on electroweak symmetry breaking. We assume that gravitational interaction is flavor blind and compute the Jarlskog determinant due to the Planck scale effects. In the case of neutrino sector, the strength of CP violation is measured by Jarlskog determinant. In this paper, we assume CP violation arise from Planck scale effects. We applied our approach to study Jarlskog determinant due to the Planck scale effects.

## I. INTRODUCTION

The origin of CP violation is still mystery in particle physics. Recent advance in neutrino physics observation mainly of astrophysical observation suggested the existence of tiny neutrino mass. The experiments and observations have shown evidences for neutrino oscillation. The Solar neutrino deficit has long been observed [1-4] the atmospheric neutrino anomaly has been found [5-7] and currently almost confirmed by KamLAND [8], and hence indicates that neutrinos are massive and there is mixing in lepton sector. Since there is a mixing in lepton sector, this indicate to imagine that there occurs CP violation in lepton sector. Several physicist have considered whether we can see CP violation effect in lepton sector through long base line neutrino oscillation experiments. The neutrino oscillation probabilities, in general depends on six parameter two independent mass square difference $\Delta_{21}$ and $\Delta_{31}$, three mixing angles $\theta_{12}, \theta_{23}, \theta_{13}$, and one CP violating phase $\delta$.

CP violation effect arise as three (or more) generation [9, 10]. CP violation in neutrino oscillation is interesting because it relates directly to CP phase parameter in the mixing matrix for n>3 degenerate neutrinos. In this paper we consider the case of three light neutrinos. We can write down the compact formula for the difference of transition probability between conjugate channel

$$\Delta P = P(\nu_\alpha - \nu_\beta) - P(\overline{\nu}_\alpha - \overline{\nu}_\beta), \qquad (1)$$

where

$$(\alpha, \beta) = (e, \mu), (\mu, \tau), (\tau, e).$$

The main physical goals in future experiments are the determination of the unknown parameter $\theta_{13}$, and upper bound $sin^2 2\theta_{13} < 0.1$ is obtained from the Ref. [11]. In particular, the observation of $\delta$ is quite interesting from the point of view that $\delta$ is related to the origin of the matter in the universe. The determination of $\delta$ is the final goal in the future neutrino oscillation experiments. We get analytical expression for $\Delta P$ and $Acp$ using the usual form of the MNS matrix parametrization [12].

$$U = \begin{pmatrix} c_{12}c_{13} & s_{12}c_{13} & s_{13}e^{-i\delta} \\ -s_{12}c_{23} - c_{12}s_{23}s_{13}e^{i\delta} & c_{12}c_{23} - s_{12}s_{23}s_{13}e^{i\delta} & s_{23}c_{13} \\ s_{12}s_{23} - c_{12}c_{23}s_{13}e^{i\delta} & -c_{12}s_{23} - s_{12}s_{13}e^{i\delta} & c_{23}s_{13} \end{pmatrix}, \quad (2)$$

where c and s denoted the cosine and sine of the respective notation, thus $\Delta P$ in vacuum can be written as

$$\Delta P(\alpha, \beta) = P(\nu_\alpha \to \nu_\beta) - P(\overline{\nu}_\alpha \to \overline{\nu}_\beta) = 16 J_{cp}(sin\Delta_{21} sin\Delta_{32} sin\Delta_{31}). \quad (3)$$

Here $\alpha$ and $\beta$ denote different neutrino (or anti-neutrino) flavours,

where

$$\Delta_{ij} = 1.27 \left(\frac{\Delta_{ij}}{eV^2}\right)\left(\frac{L}{Km}\right)\left(\frac{1GeV}{E}\right), \quad (4)$$

$\Delta_{ij} = (m_i^2 - m_j^2)$, is the difference of $i^{th}$ and $j^{th}$ vacuum mass-square eigenvalues, E is the neutrino energy, and L is the travels distance and the well known Jariskog-invariance [13] $J$ in the standard mixing parameterization is given by

$$J = Im\{U_{MNSe1} U^*_{MNSe2} U^*_{MNS\mu1} U_{MNS\mu2}\},$$

$$= \frac{1}{8} sin2\theta_{12} sin2\theta_{13} sin2\theta_{23} cos\theta_{13} sin\delta, \quad (5)$$

and the asymmetry parameter suggested by Cabibbo [13], as an alternative to measure CP violation in the lepton sector

$$Acp = \frac{\Delta P}{P(\nu_\alpha \to \nu_\beta) + P(\overline{\nu}_\alpha \to \overline{\nu}_\beta)} \quad (6)$$

CP violation in neutrino oscillation can in principal be observed in neutrino experiments by looking at the difference of the transition probabilities between CP conjugate channel. We further assume that CP violation phase arises only at Planck scale and define by

$$U'_{e3} = sin\theta'_{13} e^{-i\delta'} \quad (7)$$

Much progress has been made towards, determining the accurate values of the three mixing angles. The purpose of this paper is to study the Planck scale effects on $J$ Jarlskog determinant. In Sec-2, the neutrino mixing angles and mass square differences above the GUT Scale, is given. In Sec-3 we give the numerical analysis and results.

## II. NEUTRINO MIXING ANGLES AND MASS SQUARE DIFFERENCES ABOVE THE GUT SCALE

To calculate the effects of this perturbation on neutrino observables. The calculation developed in an earlier paper [14]. A natural assumption is that unperturbed ($0^{th} - order$) mass matrix M is given by:

$$M = U^* diag(M_i) U^\dagger, \tag{8}$$

where, $U_{\alpha i}$ is the usual mixing matrix and $M_i$ the neutrino masses is generated by grand unified dynamics. Most of the parameters related to neutrino oscillation are known, the major expectation is given by the mixing elements $U_{e3}$. We adopt the usual parameterization:

$$|\frac{U_{e2}}{U_{e1}}| = tan\theta_{12}. \tag{9}$$

$$|\frac{U_{\mu 3}}{U_{\tau 3}}| = tan\theta_{23}. \tag{10}$$

$$|U_{e3}| = sin\theta_{13}. \tag{11}$$

In terms of the above mixing angles, the mixing matrix is written as

$$U = diag(e^{if1},\ e^{if2},\ e^{if3}) R(\theta_{23}) \Delta R(\theta_{13}) \Delta^* R(\theta_{12}) diag(e^{ia1}, e^{ia2}, 1). \tag{12}$$

The matrix $\Delta = diag(e^{\frac{i\delta}{2}}, 1, e^{\frac{-i\delta}{2}})$ contains the Dirac phase $\delta$. This leads to CP violation in neutrino oscillations. $a1$ and $a2$ are the so called Majorana phases, which affect the neutrinoless double beta decay. $f1$, $f2$ and $f3$ are usually absorbed as a part of the definition of the charge lepton field. It is possible to rotate these phases away, if the mass matrix eq(9) is the complete mass matrix. However, since we are going to add another contribution to this mass matrix, these phases of the zeroth order mass matrix can have an impact on the complete mass matrix and thus must be retained. By the same token, the Majorana phases which are usually redundant for oscillations have a dynamical role to play now. Planck scale effects will add other contributions to the mass matrix. Including the Planck scale mass terms, the mass matrix in flavour space is modified as

$$\mathbf{M} \to \mathbf{M}' = \mathbf{M} + \mu\lambda, \tag{13}$$

with $\lambda$ being a matrix whose elements are all 1 and $\mu$ is small as discussed in [15]. We treat the second term (the Planck scale mass terms) in the above equation as a perturbation to the first term (the GUT scale mass terms). The impact of the perturbation on the neutrino masses and mixing angles can be seen by forming the hermitian matrix

$$\mathbf{M}'^\dagger \mathbf{M}' = (\mathbf{M} + \mu\lambda)^\dagger (\mathbf{M} + \mu\lambda), \tag{14}$$

which is the matrix relevant for oscillation physics. To the first order in the small parameter $\mu$, the above matrix is

$$\mathbf{M}^\dagger \mathbf{M} + \mu\lambda^\dagger \mathbf{M} + \mathbf{M}^\dagger \mu\lambda. \tag{15}$$

This hermitian matrix is diagonalized by a new unitary matrix $U'$. The corresponding diagonal matrix $M'^2$, correct to first order in $\mu$, is related to the above matrix by $U'M'^2U'^\dagger$. Rewriting $\mathbf{M}$ in the above expression in terms of the diagonal matrix $M$ we get

$$U'M'^2U'^\dagger = U(M^2 + m^\dagger M + Mm)U^\dagger \tag{16}$$

where

$$m = \mu U^t \lambda U. \tag{17}$$

Here $M$ and $M'$ are the diagonal matrices with neutrino masses correct to $0^{th}$ and $1^{th}$ order in $\mu$. It is clear from eq(16) that the new mixing matrix can be written as:

$$U' = U(1 + i\delta\Theta), \tag{18}$$

where $\delta\Theta$ is a hermitian matrix that is first order in $\mu$.

From eq(16) we obtain

$$M^2 + m^\dagger M + Mm = M'^2 + [i\delta\Theta, M'^2]. \tag{19}$$

Therefore to first order in $\mu$, the mass squared difference $\Delta M_{ij}^2 = M_i^2 - M_j^2$ get modified [15, 16] as:

$$\Delta M'^2_{ij} = \Delta M_{ij}^2 + 2(M_i Re[m_{ii}] - M_j Re[m_{jj}]). \tag{20}$$

The change in the elements of the mixing matrix, which we parametrized by $\delta\Theta$, is given by

$$\delta\Theta_{ij} = \frac{iRe(m_{ij})(M_i + M_j)}{\Delta M'^2_{ij}} - \frac{Im(m_{ij})(M_i - M_j)}{\Delta M'^2_{ij}}. \tag{21}$$

The above equation determines only the off diagonal elements of matrix $\delta\Theta_{ij}$. The diagonal elements of $\delta\Theta$ can be set to zero by phase invariance.

Using Eq(18), we can calculate neutrino mixing angle due to Planck scale effect,

$$|\frac{U'_{e2}}{U'_{e1}}| = tan\theta'_{12}. \tag{22}$$

$$|\frac{U'_{\mu 3}}{U'_{\tau 3}}| = tan\theta'_{23}. \tag{23}$$

$$|U'_{e3}| = sin\theta'_{13}. \tag{24}$$

We rewrite the mixing angle $\theta_{12}$, $\theta_{13}$ and $\theta_{23}$ due to Planck scale effects as:

$$tan\theta'_{12} = \sqrt{\frac{(|U_{e2}|^2 - 2Im(U_{e1}\delta\Theta_{12} + U_{e3}\delta\Theta^*_{13})U^*_{e2})}{|U_{e1}|^2 - 2Im(U_{e2}\delta\Theta_{12} + U_{e3}\delta\Theta_{13})U^*_{e1})}} \tag{25}$$

$$tan\theta'_{23} = \sqrt{\frac{(|U_{\mu3}|^2 - 2Im(U_{\mu1}\delta\Theta_{12} + U_{\mu3}\delta\Theta^*_{23})U^*_{\mu3})}{|U_{\tau3}|^2 - 2Im(U_{\tau1}\delta\Theta_{13} + U_{\tau2}\delta\Theta_{23})U_{\tau3})}} \quad (26)$$

$$sin\theta'_{13} = \sqrt{|U_{e3}|^2 - 2Im(U_{e1}\delta\Theta_{13} + U_{e2}\delta\Theta_{23})U^*_{e3}}. \quad (27)$$

As one can see from the above expression of mixing angle due to Planck scale effect, depends on new contribution of mixing angle eq(21). To see the mixing angle at Planck scale [14,16], only $\theta_{13}$ and $\theta_{12}$ mixing angle have reasonable deviation due to Planck scale effects.

### III. JARSKLOG DETERMINANT IN PLANCK SCALE :A NUMERICAL ANALYSIS

Note from eq(21), that the correction term depends crucially on the type of neutrino mass spectrum. For a hierarchical or invert hierarchal spectrum the correction is negligible. Hence we consider a degenerate neutrino spectrum and take the common neutrino mass to 2eV, which is the upper limit from the tritium beta decay experiment [17].

It is well known, CP violation at low energy responsible for non zero value of CP phase $\delta_{cp}$. At high energy (Planck scale), let us compute Jarsklog determinant due to new mixing matrix above GUT Scale

$$J_{Planck} = Im\{U_{MNSe1}U^*_{MNSe2}U^*_{MNS\mu1}U_{MNS\mu2}\}_{planck},$$

$$= Im(U_{e1} + i(U_{e2}\delta\Theta^*_{12} + U_{e3}\delta\Theta_{13})(U^*_{e2} - i(U^*_{e1}\delta\Theta^*_{12} + U^*_{e3}\delta\Theta_{23}))$$

$$(U^*_{\mu1} - i(U^*_{\mu2}\delta\Theta_{12} + U^*_{\mu3}\delta\Theta_{13})(U_{\mu2} + i(U_{\mu1}\delta\Theta_{12} + U_{\mu3}\delta\Theta^*_{23}) \quad (28)$$

which is proportional to the CP asymmetry in neutrino oscillation,

$$\Delta P = 16J \ (sin\Delta_{21}sin\Delta_{32}sin\Delta_{31}). \quad (29)$$

By using Eq(28), we simplified Jarsklog determinant due to new mixing matrix

$$J_{Planck} = \frac{1}{8}sin2\theta_{12}sin2\theta_{13}sin2\theta_{23}c + Im(i(U^*_{\mu1}U_{\mu2})(|U_{e2}|^2\delta\Theta^*_{12} + U^*_{e2}U_{e3}\delta\Theta_{13}$$

| $\delta$ | $J^*$ | $\delta$ | $J^*$ |
|---|---|---|---|
| $5^o$ | $1.06 \times 10^{-4}$ | $50^o$ | $6.10 \times 10^{-4}$ |
| $10^o$ | $2.10 \times 10^{-4}$ | $55^o$ | $5.83 \times 10^{-4}$ |
| $15^o$ | $3.07 \times 10^{-4}$ | $60^o$ | $5.37 \times 10^{-4}$ |
| $20^o$ | $3.95 \times 10^{-4}$ | $65^o$ | $4.76 \times 10^{-4}$ |
| $25^o$ | $4.71 \times 10^{-4}$ | $70^o$ | $4.00 \times 10^{-4}$ |
| $30^o$ | $5.34 \times 10^{-4}$ | $75^o$ | $3.11 \times 10^{-4}$ |
| $35^o$ | $5.80 \times 10^{-4}$ | $80^o$ | $2.13 \times 10^{-4}$ |
| $40^o$ | $6.08 \times 10^{-4}$ | $85^o$ | $1.08 \times 10^{-4}$ |
| $45^o$ | $6.18 \times 10^{-4}$ | $90^o$ | $2.21 \times 10^{-4}$ |

Table I: The unabsorbed CP phase and Jarlskog determinant. Input values are $\theta_{12} = \theta_{23} = 45^o$, $\theta_{13} = 0^o$.

$$-|U_{e1}|^2 \delta\Theta^*_{12} - U_{e1}U^*_{e3}\delta\Theta^*_{23}) + Im(i(U^*_{e1}U_{e2})(|U_{\mu 1}|^2\delta\Theta_{12} + U^*_{\mu 1}U_{\mu 3}\delta\Theta^*_{23} - |U_{\mu 2}|^2\delta\Theta_{12} - U_{\mu 2}U^*_{\mu 3}\delta\Theta_{13}),$$

$$= J_{GUT\ Scale} + J^* \quad (30)$$

From the expression of $J_{planck}$ in eq(30), it is obvious that all three CP phase and mixing angles contributes to CP violation in neutrino oscillation at Planck scale. The unabsorbed CP phase, corresponding value of $J^*$ given in Table(1). In fact, one can say that if $\theta_{12}$ and $\theta_{13}$ deviated in Planck scale region, this can be an indication of new contribution of $J^*$ due to Planck scale.

Let us define the percentage change in Jarlskog determinant due to mixing angles in Planck scale.

$$P = \frac{J^*}{J_{GUT}} \times 100 \quad (31)$$

The Majorana phases $a_1$ and $a_2$ have a non-trival effect on the Planck scale corrections. We show the results as contour plots of $J^*$ in the $a_1 - a_2$ plane. In our calculation, we used best fit values of mixing angles, $\theta_{12} = 34^o$, $\theta_{23} = 45^o$, $\theta_{13} = 10^o$. We considered non zero value of CP phase and we took $\delta = 45^o$, $90^o$.

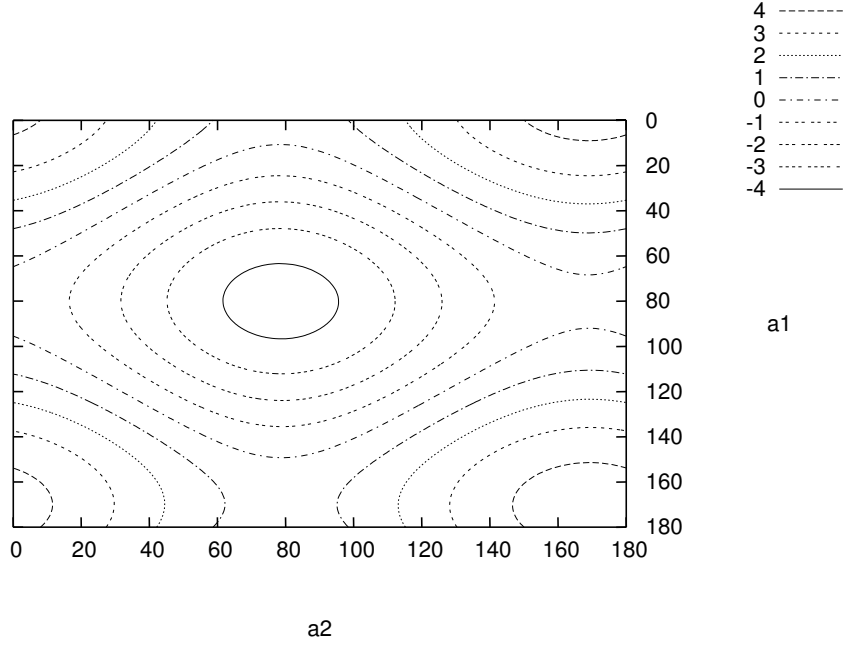

Figure 1: $J^*$ due to Planck scale as a function of Majorana phase for $\delta_{cp} = 45^o$.

In Fig. 2 and Fig. 3 for $\theta_{13} = 10^o$. For the value, which is the upper limit coming for CHOOZ experiments, note that there is reasonable range of Majarona phases, where Jarslkog determent change 5% only, this change due to only two mixing angles.

We have studied, how Planck scale effects the Jarlskog determinant for neutrino sector. MNS matrix and Jarlskog determinant, which is signal for CP violation in neutrino oscillation. We have obtained, in Planck scale, two mixing angle $\theta_{13}$, $\theta_{12}$ extra contributes to Jarlskog determinant. In present paper, we study the Jarlskog determinant due to Planck scale. We estimate the if one Dirac CP phase is responsible for CP violation at low energy then at Planck scale two mixing angle $\theta_{13}$, $\theta_{12}$ will extra contribute to Jarlskog determinant. In conclusion, we have demonstrated that two mixing angles will effect the size of Jarlskog determinant in Planck scale. In our prediction of Jarlskog determinant, two mixing angle

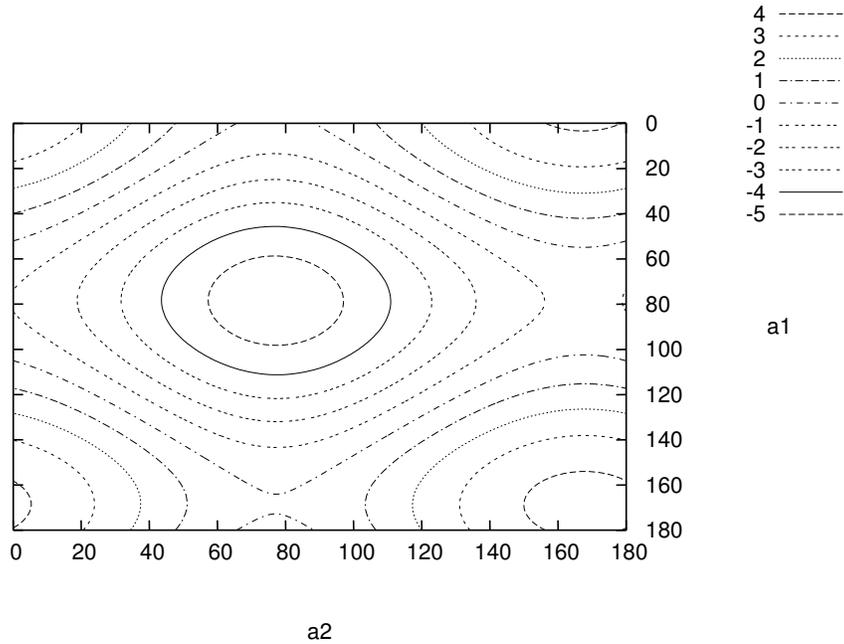

Figure 2: $J^*$ due to Planck scale as a function of Majorana phase for $\delta_{cp} = 90^o$.

$\theta_{13}$, $\theta_{12}$ will extra contributes to Jarlskog determinant in Planck scale region.

---

[1] GALLEX Collaboration, W. Hampal *et al.*, Phys. Lett **B 447**, 127 (1999).

[2] SAGE Collaboration, J. N. Abdurashitor *et al.*, astro-ph/9907113.

[3] KamioKande Collaboration, Y. Fukuda *et al.*, Phys. Rev. Lett **82**, 1810 (1999).

[4] Homestake Collaboration, B. T. Clevel and *et al.*, Astrophys. **J. 496**, 505 (1998).

[5] KamioKande Collaboration, K. S. Hirata *et al.*, Phys. Lett. **B 205**, 416(1998).

[6] IMB Collaboration, D. Casper *et al.*,, Phys. Rev. Lett **66**, 2561 (1991).

[7] MACRO Collaboration, M. Ambrosio *et al.*,Phys. Lett **B 434**, 451 (1998)

[8] KamLAND Collaboration, K. E. Guchi *et al.*, arXiv:hep-ph/021202.

[9] M. Kobayashi and T. Maskawa, Prog. Theor. Phys. Rev. Lett. **45**, 652 (1973).

[10] V. Barger, K. Whisnant and R. J. N. Phillips, Phy. Rev. lett **45**, 2084 (1980).

[11] CHOOZ Collaboration, M. Apollonio Phy. Lett **B 420**, 397 (1998).

[12] Review of Particle Physics, J. of Phys. **G 33,** 156 (2006).

[13] N. Cabibbo, Phys. lett. **B72**, 333 (1978).

[14] F.Vissani, M .Narayan and V. Berezinsky, Phys. Lett. **B 571**, 209-216, 2003.

[15] Bipin Singh Koranga, Mohan Narayan and S. Uma Sankar, arXiv:hep-ph/0607274.

[16] Bipin Singh Koranga, Mohan Narayan and S. Uma Sankar, arXiv:hep-ph/0611186.

[17] Ch. Weinneimer *et al.*, Phys, Lett. **B 460**, 219 (1999) ; V. M Labashev *et al.*, Phys. Lett. **B 460**, 227 (1999).